# Specific heat and heat conductivity of the BaTiO$_3$ polycrystalline films with the thickness in the range 20 – 1100 nm

B.A.Strukov, S.T.Davitadze, S.N.Kravchun, S.A.Taraskin,

B.M.Goltzman[1], V.V.Lemanov[1] and S.G.Shulman[1]

Lomonosov Moscow State University, Moscow 119992, Russia

[1]Ioffe Physico-Technical Institution RAS, St.-Petersburg 194021, Russia

*e-mail: bstrukov@mail.ru*

**Abstract**. Thermal properties – specific heat and heat conductivity coefficient - of polycrystalline BaTiO$_3$ films on massive substrates were studied as a function of the temperature and the film thickness by ac-hot probe method. The anomalies of specific heat with decreasing of the film thickness from 1100 to 20 nm revealed the reducing of T$_c$ and excess entropy of the ferroelectric phase transition, which becomes diffused. The critical thickness of the film at which T$_c$ = 0 estimated as 2.5 nm.

## 1. INTRODUCTION

The physical properties of ferroelectric thin films are of growing interest because of their promising applications. The reduced spatial dimensions up to nanometer scale results in the sufficient change of ferroelectric films and fine particles behavior in comparison with bulk objects. Finite-size effects are under consideration in a literature for a long time but only at the last decade the improved technology and deposition techniques provide a real possibility for obtaining of the films and particles with the controlled and reproduced parameters. It seems to be proved experimentally that the ferroelectric properties of the fine PbTiO$_3$ and BaTiO$_3$ nanometer powder disappear at the room temperature for the "critical size" of the particles that is about 10 nm for PbTiO$_3$ and 40 nm for BaTiO$_3$ [1-5]. Phenomenological analysis based on the Landau-Ginzburg theory of the structural phase transitions where the gradient and surface terms are taken into account was successively applied for the analysis of the phenomenon [6-8].

This theoretical approach is based on the milestone work [9] where the finite-size effects were considered for the magnetics in the similar way.

As to the size effects in the thin and ultrathin ferroelectric films the understanding of the associated effects are still far from the full completeness. The problem of the reliable prognosis of the properties of thin ferroelectric films is much more intricate in comparison with the case of fine particles. It is extremely hard to deal with the free-standing thin film in experiment. It means that the supporting substrate has always to be taken into account. Therefore the application of the mentioned above theory to the thin films gives result, which is in difference with experiment [7-8, 10-11].

It is widely accepted up-to-date that the real situation in the epitaxial and polycrystalline thin films can be considered only by taking into account the interaction between the film and substrate which is believed to be of the mechanical origin and results from the two-dimensional misfit stresses in the plane contact film-substrate [12-15]. These misfit strains may crucially affect the ferroelectric phase transition. The phase diagram "misfit strain-transition temperature" was obtained at [13] for $BaTiO_3$ and $PbTiO_3$ epitaxial films predicting the possible change of the sequence and order of the phase transitions in comparison with the bulk crystals. Besides the sign of the shift of a ferroelectric phase transition supposed to be determined taking into account the competition of surface effects and the misfit strains [15].

The experimental situation seems to be well contradictory. The real systems which are used in experiment for the measurements of dielectric properties and the polarization reversal are in fact the complex heterostructure which consists of a substrate, electrode (metal or some conducting substances), ferroelectric film and another electrode. The partial relaxation of the misfit strains is possible due to domain and edge dislocation patterns formation in the films [14, 16-17]; as well dead layer formation near the plane film-electrode can substantially change the domain structure and effective dielectric constant of the film [18]. We could point as an example of the unsolved discrepancies the references [19-21] where quite different results for the thickness dependence of the spontaneous polarization and the transition temperature were obtained for the epitaxial $BaTiO_3$ films grown at $SrTiO_3$ substrate.

It seems worthwhile to use the measurements of the nonelectrical parameters for the characterization of the phase transitions in the thin ferroelectric films avoiding the electrodes and therefore sufficiently simplifying the system. That was a main motivation for this study; we used the thermal properties of the films (specific heat and heat conductivity coefficient) as fundamental material parameters closely related to the phase transitions.

It should be noted that the thermal properties of thin ferroelectric films practically were not investigated inspite of their evident informativity. In fact we can indirectly estimate spontaneous polarization value through the specific heat temperature dependence as well as Curie temperature itself. It is evidently connected with a complexity of measuring of these properties for the micron and submicron films deposited on a massive substrate. For example the traditional methods of ac-calorimetry are the best for the measurements of a specific heat of samples with the thickness about 100 μm for the limited frequency near 1 Hz [22]. So far as the film is supposed to be deposited on the substrate of such a thickness only the total thermal parameters of the film-substrate structure can be obtained [23-24]. In [24] the specific heat of $BaTiO_3$ epitaxial films with thickness 6 and 200 nm was measured with $SrTiO_3$ as substrate. No clear anomalies at the phase transitions were observed; we shall return to these results later.

## 2. EXPERIMENTAL

Recently we have shown that ac-hot probe method can be successfully applied for the thermal characterization of thin dielectric films deposited on the massive substrate [25-26]. The basic idea of the heater-probe method is that a metal strip or wire made of a metal with high thermal resistivity coefficient serves as a source of heat and a temperature sensor simultaneously. The probe is heated by periodic sinusoidal current with a given frequency ω (as a rule these frequencies f = ω/2π are in Hz-kHz region). The probe temperature oscillations depend on the thermal properties of surrounding media. These oscillations can be measured using temperature dependence of the probe resistance. The temperature and resistivity oscillations occur at a frequency 2ω, and the voltage across the probe has a frequency component 3ω. Measuring the amplitude and phase of the voltage at the frequency 3ω one can determine the specific heat $c$ and the thermal conductivity λ of the layer with thickness about to the thermal wave penetration depth. It was revealed that for the condition

$$h_f \ll d_p \ll h_s$$

(here $h_f$ is the thickness of the film, $d_p$ thermal waves penetration depth ($d_p = (\lambda_s/2c_s\omega)^{1/2}$, $\lambda_s$, $c_s$ heat conductivity coefficient and specific heat of a substrate), $h_s$ the thickness of the substrate), important asympthotic relation can be obtained for the ratio of the complex temperature oscillations of the probes deposited on the film and on the substrate (Figure 1).

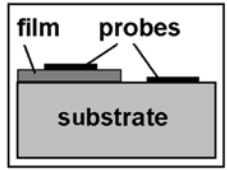

**Figure 1.** Sample arrangement: substrate, film and hot-probes positions

This ratio has a form [26]

$$\delta T_f/\delta T_s = 1 + (1 + i)(X_{fs} - 1/X_{fs})(2\omega c_f/\lambda_f)^{1/2} h_f \qquad (1)$$

where $X_{fs} = (\lambda_s c_s/\lambda_f c_f)^{1/2}$ is the thermal contrast factor (here $\lambda_f$, $c_f$ heat conductivity coefficient and specific heat of the film). It is seen that if $X_{fs}^2 \gg 1$, the Eq. (1) has the following asymptotic form:

$$\delta T_f/\delta T_s \approx 1 + (1 + i)(\lambda_s c_s)^{1/2} h_f \omega^{1/2}/\lambda_f \qquad (2)$$

At the other limit case, if $X_{fs}^2 \ll 1$

$$\delta T_f/\delta T_s \approx 1 - (1 + i)(\lambda_s c_s)^{-1/2} h_f \omega^{1/2} c_f \qquad (3)$$

Therefore we found that the role of the thermal contrast factor is very important: it depends on the $X_{fs}$ value what parameter of a film can be determined from the analysis of the probe temperature oscillation. As follows from equations (2) and (3) it is possible to determine the whole set of thermal parameters of a film, sputtered on the massive substrate by measuring the amplitudes of temperature oscillation of two probes on two different substrates which satisfy to the conditions $X_{fs}^2 \gg 1$ (measurements of the heat conductivity coefficient) or $X_{fs}^2 \ll 1$ (measurements of the specific heat). For $BaTiO_3$ thin films a fused quartz and a crystalline sapphire can be accepted as substrates satisfying the mentioned conditions.

Our experimental setup was described in [26] in details. The final accuracy of the measurements of $c_f$ and $\lambda_f$ at the temperature region 300-400 K was estimated as 3-5% and 2-3% for lower temperatures; film thickness variation was 20 – 1100 nm.

## 3. SAMPLE PREPARATION

$BaTiO_3$ films were deposited on fused $SiO_2$ and monocrystalline sapphire substrates $Al_2O_3$ (the typical size 4.0x5.0x0.5 mm) by rf magnetron sputtering (f = 13.5 MHz) with the substrate position "off axis" at 750°C substrate temperature. X-ray diffraction for the $BaTiO_3$ films on $Al_2O_3$ revealed <111> textured polycrystalline films with a tetragonal symmetry. The half-width of {110} rocking curve was 5' – 7'. A mixture of $Ar:O_2$ gas at the 0.8/0.2 ratio was introduced and the total pressure was held at $4·10^{-2}$ Torr. Deposition rate was estimated by the control of sputtering time as 5 nm/min. It is important that the $BaTiO_3$ films were deposited on the prepared and etched surface of substrate to have the necessary intimate contact film-substrate.

The surfaces of all the prepared films were inspected by the atomic force microscope. The average size of the monocrystalline grains and a roughness of the surfaces for the different films are presented in table 1.

**Table 1.** Thickness dependence of the surface roughness $\Delta h$ and of the average size D of $BaTiO_3$ crystallites for the different substrates

| Substrate | Film thickness, nm | $\Delta h$, nm | D, nm |
|---|---|---|---|
| Sapphire $Al_2O_3$ | 1400 | 24 | 176 |
| | 1100 | 21 | 168 |
| | 400 | 22 | 140 |
| | 300 | 15 | 104 |
| | 40 | 15 | 93 |
| | 20 | 10 | 25 |
| Fused quartz $SiO_2$ | 1100 | 38 | 165 |
| | 400 | 13 | 170 |
| | 300 | 15 | 105 |
| | 100 | 13 | 125 |
| | 70 | 3 | 143 |
| | 40 | 4 | 90 |
| | 20 | 8 | 42 |

Figure 2 illustrates the typical picture of the film surface with the thickness 100 nm on a quartz and sapphire substrate.

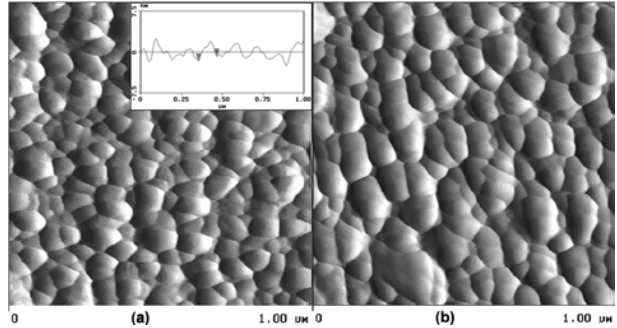

**Figure 2.** AFM image of the 100 nm BaTiO$_3$ film deposited on fused SiO$_2$ (a) and Al$_2$O$_3$ (b). Inset: roughness of the film shown at (a)

## 4. EXPERIMENTAL RESULTS AND DISCUSSION

The set of basic thermal parameters of BaTiO$_3$ films on SiO$_2$ and Al$_2$O$_3$ substrates were determined in the wide temperature interval 80-420 K both on cooling and on heating. The more detailed study has been done for the temperature region where the ferroelectric phase transition occurs in the bulk crystal (360-410 K).

The temperature dependence of c and λ for 1100 nm film is presented at Figure 3.

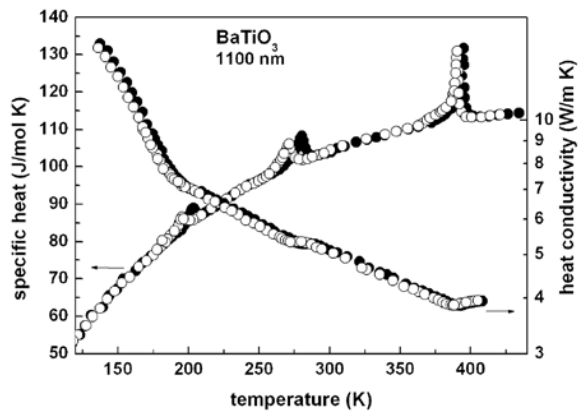

**Figure 3.** Temperature dependences of the specific heat and heat conductivity coefficient for BaTiO$_3$ 1100 nm film on heating (•) and on cooling (o).

All three phase transitions typical for bulk BaTiO$_3$ are represented clearly: $m3m \rightarrow 4mm \rightarrow mm2 \rightarrow 3m$ for temperatures 395, 280, 204 K (on heating) respectively. The temperature hysteresis for all transitions was detected for c(T) dependence revealing the main first order phase transitions feature. It worth to note that the excess energy $\Delta Q$ of the ferroelectric phase transition was found to be 150 J/mol, that is somewhat lower than the value obtained earlier for the bulk ceramic (190 J/mol in [28]). The temperature dependence of the heat conductivity coefficient for this thick film is shown at Figure 3 and seems to be similar to one for a bulk crystal [29]. Therefore these films can be considered as "thick" from this point of view.

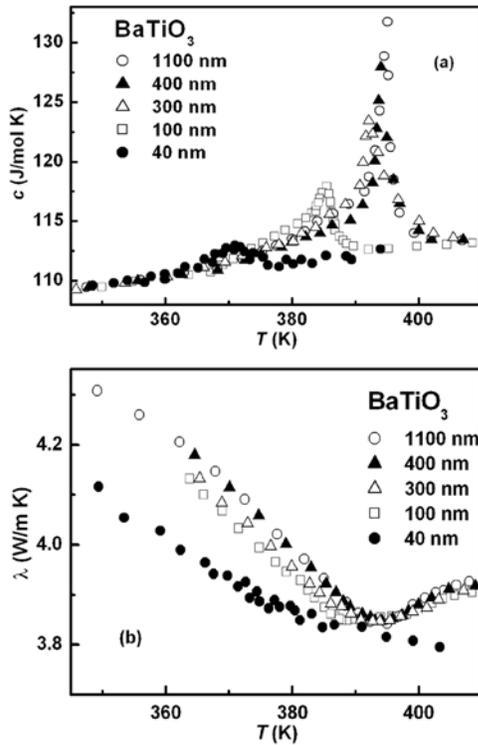

**Figure 4.** Temperature dependences of the specific heat (a) and the heat conductivity coefficient (b) in the region of ferroelectric phase transition in BaTiO$_3$ films with different thickness

The same measurements were performed for thinner films. Earlier we have presented the data for the films with the thickness higher than 0.1 µm [30]. Now this range was expanded down to 20 nm. The results are presented in Figures 4a and 4b. It should be noted that no anomalies were detected for 20 nm film and these data are not shown here. It is clear that when the thickness of the film reduces, the lowering of the phase transition temperature occurs accompanied by the increasing diffusion of the anomalies. The temperature hysteresis of $T_c$ of about 5K was marked for the films 100 nm and thicker revealing a preservation of the first order transition. No hysteresis was observed at 40 nm film, that is probably because of the large diffusion of the transition. In fact the anomaly of $\lambda$ is very weak for this film as well; it is possible that this effect is connected with the increasing of the film roughness for the submicron films [31].

The observed anomalies of the specific heat for the films under consideration can be used for the determination of the transition temperature and the excess entropy change;

besides the temperature dependence of a spontaneous polarization can be calculated with the help of the simple thermodynamic relations [32].

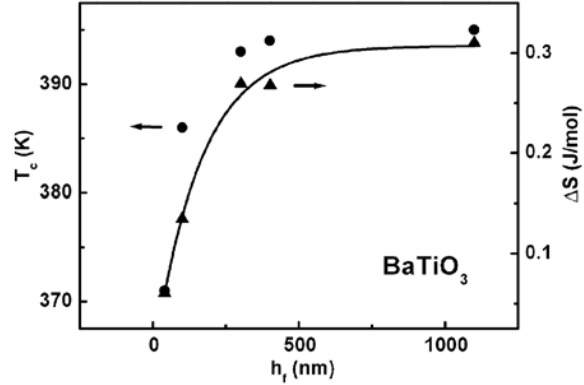

**Figure 5.** Temperature and the excess entropy of the ferroelectric phase transition in BaTiO$_3$ films as a function of thickness. Line is plotted as guide to eye.

Figure 5 illustrate the thickness dependence of the transition temperatures (which were determined as temperatures of the maximal value of the specific heat) and of the excess entropy of the ferroelectric transition in BaTiO$_3$ films; it is seen that the both parameters have the clear tendency to decrease with film thickness decreasing. The quantitative values are gathered in the table 2.

**Table 2.** Thickness dependence of the transition temperature $T_c$, excess entropy and spontaneous polarization $P_s$ for BaTiO$_3$ films

| Film thickness, nm | $T_c$, K | $\Delta S$, J/(mol K) | $P_s$, µC/cm$^2$ |
|---|---|---|---|
| 1100 | 395 | 0.31 | 11 |
| 400 | 394 | 0.26 | 10 |
| 300 | 393 | 0.27 | 10 |
| 100 | 387 | 0.14 | 7.5 |
| 40 | 371 | 0.06 | 4 |

As it was pointed earlier no anomalies were observed at the film with $h_f = 20$ nm. The temperature dependences of the spontaneous polarization restored from the specific heat data are presented in Figure 6. It is evident from Figure 6 the progressive suppression of the spontaneous polarization of the films with the reducing of their thickness.

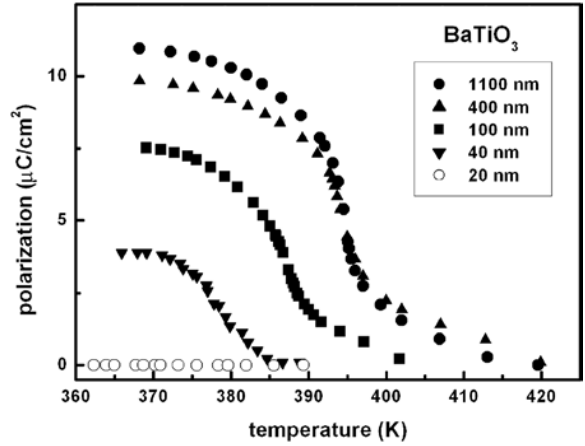

**Figure 6.** Temperature dependence of the spontaneous polarization in BaTiO$_3$ films restored from the specific heat data

Considering all the data one have to take into account that there are two dimensional parameters in a polycrystalline film: thickness of the film and the average size of crystalline grains. The question is what the parameter determines the observed evolution of the film properties. It follows from the table 1 that the average size of crystallites is practically the same for the film thickness down to 100 nm; it means, first of all, that the polycrystalline system under consideration has a columnar structure; so we can suppose that just the film thickness is the determining parameter. Evidently we have here no essential effect of the two-dimensional stresses pointed above. It is difficult to reveal exactly what is the reason for this relaxation of the film-substrate interaction in our case. Nevertheless it is clear that the main features of the thickness evolution are very close to those predicted for the free-standing films or fine particles. It is interesting to note that the dependence of the transition temperature on the reciprocal thickness of the film is close to linear (Figure 7). Such a behavior was predicted earlier for fine particles [6]; besides it was shown in [14] that the similar dependence can be obtained for the films of the proper ferroelastic-ferroelectric where the relaxation of the surface stresses is connected with the domain formation. The extrapolation of the obtained dependence $T_c = T_c^{bulk} - 1000/h_f$ with $T_c^{bulk}$ = 396 K to $T_c$ = 0 gives $h_f \approx$ 2.5 nm as a "critical thickness" for the complete suppressing of the phase transition in BaTiO$_3$ polycrystalline films.

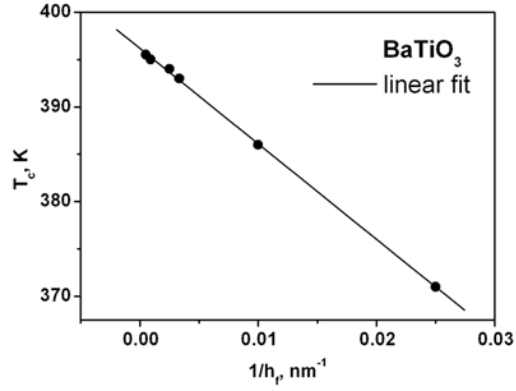

**Figure 7.** Dependence of the transition temperature on the reciprocal thickness of the BaTiO$_3$ films

It is worth to note that the similar value of the "critical thickness" (2 nm) was obtained in [34] for BaTiO$_3$ by the atomic-level simulation approach.

Returning to the reference [24] we should note that the thin epitaxial films of BaTiO$_3$ on SrTiO$_3$ substrate studied in this work showed quite different behavior of the specific heat even for $h_f$ = 200 nm. It seems that in this case the two-dimensional clamping of the film by the substrate is the determining factor contrary to our case.

## 5. CONCLUSIONS

In conclusion, we have presented data, which give the full characterization of the thermal properties of polycrystalline BaTiO$_3$ films deposited on the massive SiO$_2$-glass and Al$_2$O$_3$ substrates. The progressive evolution of the specific heat and the heat conductivity coefficient anomalies provided the unique information about finite-size effects in these films revealing the quite limited influence of the film-substrate interface mechanical stresses. The critical thickness of the films can be estimated from the data. It is supposed that the study of ferroelectric thin films by means of specific heat measurements gives an additional tool for the analysis of their properties and phase transitions.

**6. ACKNOWLEDGEMENTS**

Authors wish to express their gratitude to the Russian Foundation for Basic Research (Project 00-02-16916) and Program "Universities of Russia" for the financial support.

**REFERENCES**

[1] Uchino K, Sadanaga E and Hurose T 1989 *J. Am. Ceram. Soc.* **72** 1555

[2] Zhong W L, Jiang B, Zhang P L, Ma J M, Cheng H M, Yang Z H and Lu L X 1993 *J. Phys.: Condens. Matter* **5** 2619

[3] Chattopadhyay S, Ayyub P, Palkar V R and Multani M 1995 *Phys.Rev.B* **52** 13177

[4] Wang C L and Smith S R P 1995 *J. Phys.: Condens. Matter* **7** 7163

[5] Ishikawa K and Uemori T 1999 *Phys. Rev. B* **60** 11841

[6] Zhong W L, Wang Y G, Zhang P.L and Qu B D 1994 *Phys. Rev. B* **50** 689

[7] Li S, Eastman J A, Vetrone J M, Foster C M., Newnham R E and Cross L E 1997 *Jpn. J. Appl. Phys.* **36** 5169

[8] Ishibashi Y, Orihara H and Tilley D R 1998 *J. Phys. Soc.Jap.* **67** 3292

[9] Kaganov M I and Omelyanchouk A N 1976 *Sov. Phys. JETF* **34** 895

[10] Wang Y G, Zhong W L and Zhang P L 1996 *Phys. Rev. B* **53** 11439

[11] Zhang J, Yin Z, Zhang M C and J F Scott 2001 *Solid State Commun.* **118** 241

[12] Speck J S and Pompe W 1994 *J. Appl. Phys.* **76** 466

[13] Pertsev N A , Zembilgotov A G and Tagantsev A K 1998 *Phys. Rev. Lett.* **80** 1988

[14] Bratkovsky A M and Levanyuk A P 2001 *Phys. Rev. Lett.* **86** 3642

[15] Zembilgotov A G, Pertsev N A, Kohlstedt H and Waser R 2002 *J. Appl. Phys.* **91** 2247

[16] Speck J S, Seifert A, Pompe W and Ramesh R 1994 *J. Appl. Phys.* **76** 477

[17] Alpay S P and Roytburd A L 1998 *J. Appl. Phys.* **83** 4714

[18] Bratkovsky A M and Levanyuk A P 2000 *Phys. Rev.Lett.* **84** 3177

[19] Terauchi H, Watanabe Y, Kasatani H, Kamigaki K, Yano Y, Terashima T and Bando Y 1992 *J. Phys. Soc. Jap.* **61** 2194

[20] Yoneda Y., Okabe T, Sakaue K and Terauchi H 1998 *J. Appl. Phys.* **83** 2458

[21] Yanase N, Abe K., Fukushima N and Kawakubo T 1999 *Jpn. J. Appl. Phys.* **38** 5305

[22] Sullivan P F and Siedel G 1968 *Phys. Rev.* **173** 679

[23] Fominaga F, Fournier T, Gandit P and Chaussty 1997 *Rev. Sci. Instr.* **68** 4191

[24] Onodera A, Kawamura Y, Okabe T and Terauchi H. 1999 *J. Europ. Ceram. Soc.* **19** 1477

[25] Kravchun S N, Davitadze S T, Mizina N S and Strukov B A 1997 *Phys. Solid State* **39** 675

[26] Davitadze S T, Kravchun S N, Mizina N S, Strukov B A, Goltzman B M, Lemanov V V and Shulman S.G. 1998 *Ferroelectrics* **208-209** 279

[27] Kelman M B, Schloss L F, Intyre P C M Hendrix B C, Bilodeau S M and Roeder J F 2002 *Appl. Phys. Lett.* **80** 1258

[28] Volger J 1952 *Philips Res. Rep.* **7** 21

[29] Mantle A J H and Volger J 1967 *Phys. Lett.* **24A** 137

[30] Davitadze S T, Kravchun S N, Strukov B A, Goltzman B M, Lemanov V V and Shulman S G 2002 *Appl. Phys. Lett.* **80** 1631

[31] Lu H and Chu J 2001 *Phys. Stat. Sol.* **222** 35

[32] Strukov B A and Levanyuk A P 1998 *Ferroelectric Phenomena in Crystals* (Heidelberg: Springer) p 303

[33] Tinte S and Stachotti M G 2001 *Phys. Rev. B* **64** 235403